# ADS Standardization Landscape: Making Sense of its Status and of the Associated Research Questions

Scott Schnelle & Francesca M. Favaro, Waymo LLC

## Abstract

Automated Driving Systems (ADS) hold great potential to increase safety, mobility, and equity. However, without public acceptance, none of these promises can be fulfilled. To engender public trust, many entities in the ADS community participate in standards development organizations (SDOs) with the goal of enhancing safety for the entire industry through a collaborative approach. The breadth and depth of the ADS safety standardization landscape is vast and constantly changing, as often is the case for novel technologies in rapid evolution. The pace of development of the ADS industry makes it hard for the public and interested parties to keep track of ongoing SDO efforts, including the topics touched by each standard and the committees addressing each topic, as well as make sense of the wealth of documentation produced. Therefore, the authors present here a simplified framework for abstracting and organizing the current landscape of ADS safety standards into high-level, long term themes. This framework is then utilized to develop and organize associated research questions that haven't yet reached widely adopted industry positions, along with identifying potential gaps where further research and standardization is needed.

## Introduction

Automated Vehicles (AVs), specifically those equipped with SAE Level 4 Automated Driving Systems (ADS), have the potential, and have already started to show the capacity, for improvements in roadway safety, mobility, and equity as a part of a holistic Safe System approach (NHTSA, 2022) (FHWA, 2023) (GHSA, 2022) (Johns Hopkins University Bloomberg School of Public Health, 2021). However, a key part of ADSs achieving their potential is public acceptance. Reaching this public acceptance by engendering public trust needs to be a collaborative industry effort, not done in isolation by any one entity. To this end, many ADS developers, OEMs, suppliers, academics, government agencies, and safety advocates engage in standards development organizations (SDOs) that aim to collaboratively enhance uniformity of terminology, safety, compatibility, interoperability, and quality for the entire industry. These standardization organizations and their constituent committees and subcommittees span a breadth of topics, from those addressing a more holistic picture of transportation's systems safety to those addressing ADS specific issues. The structure of these standards organizations differ too, with some specializing in ADS-dedicated committees, such as the SAE On-Road Automated Driving (ORAD)[1] with 10 active Task Forces within it, and others subsuming ADSs under broader categories, such as the ISO TC 22 SC 32 on Road Vehicles Electronics and Electrical (EE) Subsystems[2], where the ISO 5803 technical report on ADS safety (ISO/AWI TS 5083) is being drafted by a newly developed sub-working group with ADS technology becoming an instantiation of electronics and electrical automotive components. Beyond the breadth of topics and themes within these committees, there is also a wide range in the level of technicality and detail intrinsic to the nature of each group and standard. For example, documents such as ISO 5083 are more foundational in content, while others like IEEE 2846-2022 (IEEE, 2022) are more technical. Despite their differences, all these efforts aim at providing consensus standards to guide this nascent industry and foster public acceptance. These standards additionally have the potential to provide useful information towards the development of appropriate regulatory frameworks that permit the safe operation of ADS on public roads. As such, work happening in SDOs is instrumental in achieving the promises of ADS technology.

Given the breadth and depth of standards activities mentioned, it is becoming increasingly difficult to keep track of which organizations, committees, and standards are addressing what topics. This is exacerbated by the rapid evolution of the external landscape over the last two years in terms of determining what constitutes a safe ADS. Moreover, all of these SDOs are constantly producing novel documentation to try and keep up with the rapid progress of ADS technology. With the multitude of standards activities, their disparate structures, the breadth and depth of topics covered, along with the volume of documents being produced, a number of efforts exist that attempt to provide tools to track progress of these organizations and standards. Two examples of such efforts include the SAE[3] and BSI roadmap[4] tools. In addition to these tools, liaison agreements and memorandum of understanding (MOUs) at the standards committee levels ensure cross-pollination and collaboration within and across SDOs.

Despite such collaborative efforts, developing a coherent organization and overview of these efforts and the documentation they produce is becoming exceedingly hard and any effort that attempts to stay up to date quickly becomes antiquated. Therefore, to help organize and disentangle the landscape of ADS safety standards, this paper abstracts high level and long term themes from these SDOs and their work. The goal is to bring forth a mapping of specific topics of interest to broader, but flexible, categories. To help focus this effort, the scope of the high level categories provided in this framework are restricted to the safe development and deployment of ADSs, mainly trying to answer the central question: "how safe is safe enough?" for the specific case of ADS applications. To provide context, some specific examples of standards and external references are provided in Table 1, however, for the reasons previously mentioned, this list is not meant to be exhaustive nor will it remain up-to-date for long.

Beyond providing a structure for organizing the ADS standardization landscape, the framework presented in this paper also provides associated research questions for each high-level category, which we think still haven't reached full industry consensus. In fact, due to the rapid progress in the development and deployment of Level 4 ADSs, many standardization efforts are essentially geared towards collaborative *research and understanding* rather than documentation of a standardized and mature know-how. With only a few companies fielding Level 4 ADS to date, the type of collaboration and debate happening in relation to ADS safety standards

---

development and safety determination is reflective of this more investigative nature rather than solidification and standardization of information and practices. This affects, in turn, the type of documents developed so far, with the majority of them addressing broader, more foundational content such as definitions and terminology, and many remaining anchored in technical and information reports than more normative counterparts[5]. These foundational standards ensure that the same type of lexicon and vocabulary exists to facilitate a shared understanding which can then enable the discussion of more in-depth processes and technical issues. This is important as the global interest in ADS continues to increase, with potential asynchronous attention from different countries and interest groups, each focusing on varying aspects of automation. Overall, the industry is in need of global harmonization of both vocabulary and processes so that the current focus of the vast majority of SDO groups is well justified.

## ADS Safety Standards Organization Framework

As stated, the desired outcome of this paper is to provide structure and inform the understanding of this complex and rapidly evolving ADS standards landscape. The structure proposed here may also help identify outstanding research questions that deserve attention in these fora. To enable an abstraction of such themes and create a cohesive mapping, the authors leverage their personal experiences in a number of external engagement contexts, including:

- Engagement and leadership in external standardization activities (e.g., IEEE, SAE and ISO), both at the domestic as well as the international level
- Engagement in state of the art research, most often tied back to: i) internal research from current and past employment, ii) external engagement in public-facing research boards (e.g., Transportation Research Board[6] ); iii) participation in state-of-the-art conferences/workshops/events; and iv) service in the role of peer-reviewers within the broader academic community.
- Engagement in policy-oriented work, through participation in trade associations and industry consortia, also at both the domestic and international level.
- Engagement in commenting on regulatory drafts proposed for public consultation, again at both at the domestic and international level.

From this experience, the authors have created the ADS Safety Standards Organization Wheel presented in Figure 1,  aimed at abstracting and organizing high level and long term themes for ADS SDOs.

---

[5] As an example, the extremely well known SAE J3016 document (SAE, 2021) was first published in 2014 as an Information Report, being upgraded to Recommended Practice after over 2 years of wide-spread adoption of the Levels of Automation taxonomy included therein (the document is much richer than such a taxonomy, albeit broader audiences are often unaware of such additional content). Revisions to its content are happening at regular cadence, now in conjunction with the joint ISO Publicly Available Specification PAS 22736:2021 (ISO, 2021b).
[6] https://www.nationalacademies.org/trb/transportation-research-board

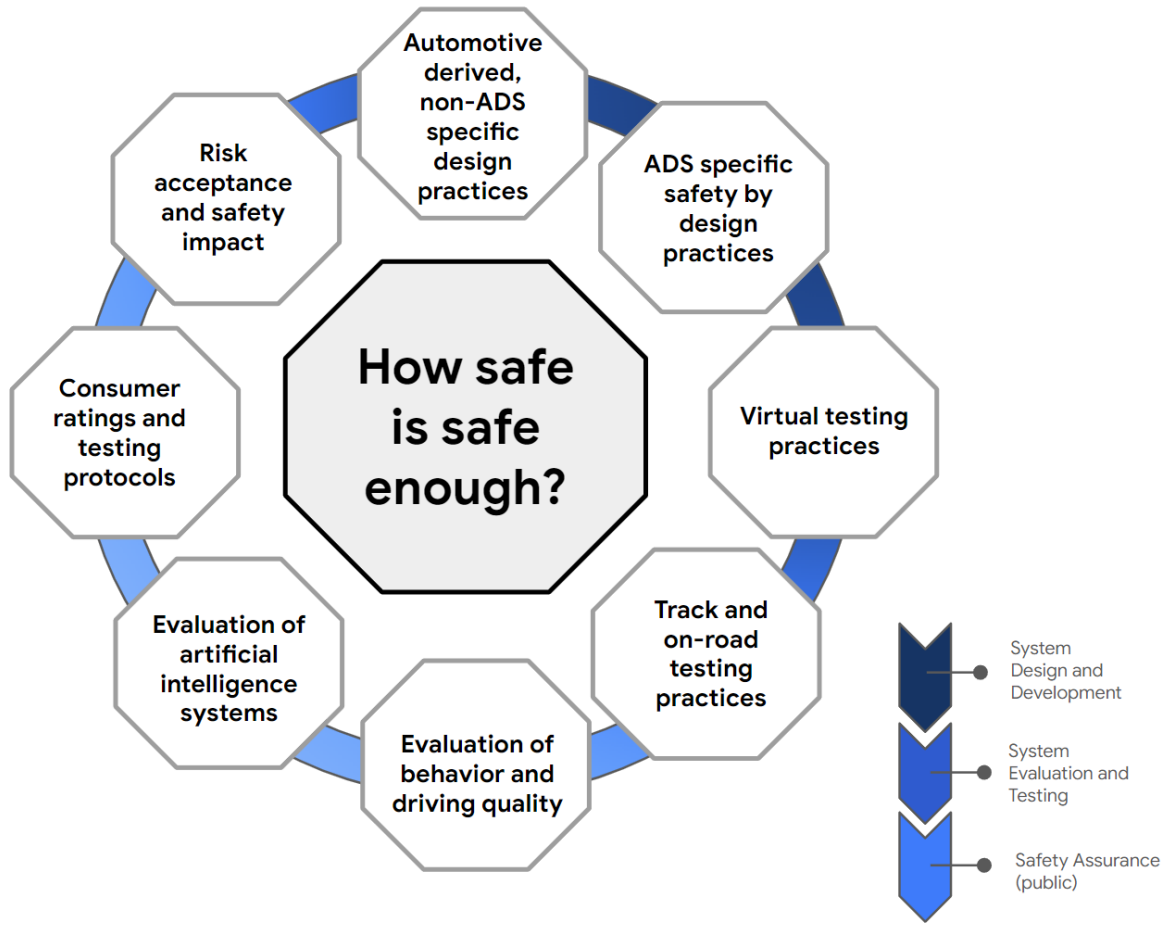

Figure 1: ADS Safety Standards Organization Wheel

The ADS Safety Standards Organization Wheel addresses standards and themes that go beyond the aforementioned first wave of foundational standards (e.g., terminology, definitions, etc.) to explore themes and standards pertinent to the question at the center of the wheel, and at the center of the safety standardization debate, of "How safe is safe enough?". The wheel symbolizes how, just like it is not possible to pinpoint a single metric to answer that question, it is not possible to look at a single standard related to the design approach or evaluation process to answer the question. Instead, all of the building blocks of the wheel, which are complex and multifaceted, are required in order to answer the central question. The eight themes represented in the wheel are as follows:

1. Automotive derived non-ADS specific design practices
   ○ Standards and practices related to the traditional automotive domain that are applicable[7] but not specific to ADS technology
2. ADS-specific safety by design processes

---

[7] Applicability is here intended broadly, and may require adaptation of certain concepts for the specific use-case of ADS technology

- Standards and practices directly related to ADS-application specific implementations of traditional safety-by-design development practices
3. Virtual testing practices
   - Standards and practices related to ADS verification and validation activities utilizing simulation
4. Track and on-road testing practices
   - Standards and practices related to ADS verification and validation activities utilizing closed-course testing and on-road testing
5. Behavioral evaluation and driving quality
   - Standards and practices related to the evaluation of the ADS behavior and its driving quality, including drivership
6. Evaluation of AI systems
   - Standards and practices related to the evaluation of artificial intelligence as it is used in ADS development and safety assurance
7. Consumer rating and testing protocols
   - Standards and practices related to ADS consumer ratings and testing protocols to help engender public trust
8. Risk acceptance and safety impact
   - Standards and practices related to evaluating the safety impact/benefits stemming from ADS deployment and related frameworks for risk acceptance

Within Figure 1, these eight building blocks are further overlaid on a circle containing various shades of blue. This is intended to represent each block's connection to the ADS development and safety determination lifecycle, with specific phases on system design, evaluation and testing, and safety assurance.

## ADS Safety Standards Open Research Questions

In addition to the ADS Safety Standards Organization Wheel, the authors provide some associated research questions as shown in Table 1. These questions attempt to identify and align on topics that deserve further attention, potentially aiding in the discovery of research areas that are not fully addressed by standards organizations. Non-exhaustive external references, including standards and policy documents, in these areas are provided as examples.

Table 1: ADS Safety Standards Categories and Associated Research Questions[8]

| Topical Mapping | High-Level Research Questions | External References |
|---|---|---|
| **Safety By Design Practices** *[Includes both ADS specific and non-ADS specific]* 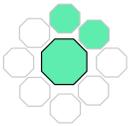 | • How do you ensure **absence of unreasonable risk (AUR)** based on **state of the art development and design practices**?<br>• How do we balance **applicability** and **adaptability** of non-ADS specific standards? For example:<br>  ○ Is it possible to **benchmark to** state of the art automotive, but **not ADS specific, standards**? If not, which parts can we explicitly leverage and which part can you not? **Why**?<br>  ○ Can flexibility in existing concepts be increased to embed ADS applications?<br>• What high level requirements and **safety goals** should be defined for the system?<br>• How do you **systematically identify, evaluate the risk for, and mitigate all applicable hazards for an ADS service**?<br>• How do you **aggregate** multiple sources of **residual risks** into a single determination of AUR? What if the actual risks and hazards considered aren't commensurable (e.g., risk of collision vs. risk of fire)<br>• How do you **minimize exposure to hazardous events**? Can you actually compute rates of hazardous causes and triggering conditions and exposure to hazards or should you consider alternative means to estimate risk? | • ISO 26262 - Functional Safety (ISO, 2018a)<br>• ISO 21448 - Safety of the Intended Functionality (ISO, 2022a)<br>• ISO 15026 - Software Assurance (ISO, 2019)<br>• SPICE Software Development Practices from ISO 3300x[9] (ISO, 2015)<br>• NATM Guidelines for Validating ADS (GRVA, 2022)<br>• ISO 5083 - ADS Design and V&V (ISO/AWI TS 5083)<br>• SAE J3208 - Taxonomy and Definitions of ADS V&V[10] |
| **Virtual Testing Practices** | • When do you have **enough tests**? How do you define **completeness /representativeness /exhaustiveness of these tests** and how are these concepts related to each other?<br>• What **requirements** do these concepts place on a **library of scenarios**? | • ISO 3450x: Scenario-based testing vocabulary (ISO, 2022b), safety evaluation framework (ISO 2022c), ODD |
| 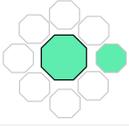 | - How do you define **reasonably foreseeable scenarios**? Does that account for **reasonably foreseeable behavior from other agents** on the road and/or also **exposure to conflict types** based on the ADS behavior?<br>- How does evaluation of reasonably foreseeable scenarios **differ** from evaluation of **edge-cases**? What specific approaches should be recommended to tackle each?<br>- How do you **parametrize the space of conflicts** and what **language** should one standardize to ensure we create **MECE sets of outcomes of interest**?<br>- How do you assess coverage against various levels of **scenario criticality** and how is **difficulty/ criticality defined**?<br>- How do you define **similarity across scenarios**?<br>- How are **ADS simulations validated**? What are elements that require validation and establish a minimum **credibility framework for realism and fidelity**?<br>- What are **standardized processes for counterfactual** simulations and logs/other events **reconstructions**? How are **alignment procedures** and **initial conditions settings standardized**, also for **comparison with performance** from artificial reference **models**?<br>- How can results be **post processed** so that an **independent reviewer** can confidently review them?<br>- How can you simulate and analyze degraded conditions? | specification[11], scenario categorization[12], and scenario evaluation and test case generation[13]<br>- SAE J3279 - Best Practice for Developing and Validating Simulation for ADS[14]<br>- NATM Guidelines for Validating ADS (GRVA, 2022)<br>- Pegasus Family: VVM[15]<br>- IAMTS virtual toolchains best practice (IAMTS, 2021)<br>- ISO 21934: Prospective safety performance assessment of pre-crash technology by virtual simulation (ISO, 2021a) |
| **Track and On-Road Testing Practices**<br><br>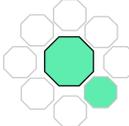 | - What are the **specific use-cases** that should be considered for on-road testing?<br>- How can closed-course testing be **efficiently** used **in conjunction** with virtual and on-road testing?<br>- What are the **best practices** for testing ADS on closed courses? What **requirements or recommendations** do they carry for testing providers and testing auditors?<br>- What are the **best practices for testing ADS on public roads?** What requirements or recommendations do they | - SAE J3018 -Guidelines for Safe On-Road Testing of ADS (SAE, 2020)<br>- PAS 1884 - Safety operators in automated vehicle testing and trialing guide (BSI, 2021) |
| | carry for ADS developers? How are they dependent on **system maturity** and **proof of performance**?<br><br>• What are **differences/similarities** in testing in closed-course compared to public road testing?<br><br>• What are the **differences/similarities** to existing testing operation and operator training guidelines? Are there any gaps? | • [SAE J3247 - ADS Test Facility Safety Practices][16]<br>• AVSC Recommended Practice for In-Vehicle Fallback Test Driver (AVSC, 2019) |
| **Evaluation of Behavior and Driving Quality** | • What is **good driving behavior**? How does one judge driving quality? How do you define proper **Drivership competencies** (Fraade-Blanar, 2022)?<br><br>• Can **behavioral envelopes** be established to assess and/or constrain **the ADS behavior**?<br><br>• How do you define reasonably foreseeable **behaviors from others**? What are appropriate **assumptions** that can be placed (on the behavior of others or of the ADS) to make such envelopes more or less stringent? How can these assumptions be **optimized and validated**?<br><br>• How do you define and quantify **predictable behavior**?<br><br>• How do you define safe and **appropriate responses** to the actions of others?<br><br>• How do you **translate the qualification of (un)reasonable behavior into a quantification of (un)reasonable risk**?<br><br>• How do you **validate models** used for behavioral references? What data do you need to provide **sufficient evidence** ( e.g., comparison of distributions for gap selection with VRUs compared to human reference from naturalistic databases)?<br><br>• What role does compliance to **traffic rules** play in good driving **behavior evaluation** for the ADS and for reasonably foreseeable actions of other road users?<br><br>• How do you derive **acceptance criteria** from a codified list of **behavioral competencies**? | • IEEE 2846 - Assumptions for Vehicle Decision-Making (IEEE, 2022)<br>• [J3237 - Operational Safety Metrics][17]<br>• AVSC Best Practice for Metrics and Methods for Assessing Safety Performance of ADS (AVSC, 2021a)<br>• EU Commission Implementing Regulation for ADS Type Approval (EU, 2022)<br>• ISO 5083 - ADS Design and V&V (ISO/AWI TS 5083) |
| **Evaluation of AI Systems**<br><br>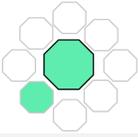 | <ul><li>Does the **development of AI systems** require more in-depth **evaluation beyond black box** approaches that test and validate the performance output?</li><li>Could safety-by-design approaches applied to AI enable sufficiency of black-box testing? And what would be ADS-specific principles for AI rigorous development?</li><li>How is **robustness** of the algorithms and **development rigor** assessed? How does explainability aid such evaluation?</li><li>How do we better discern safety-criticality for AI-enabled functions in an ADS (compared to all of them being safety-critical)?</li><li>How do **updates** in our SW **impact** our safety assessment? Is a CACE (chance anything changes everything) approach really practicable? How do updates trigger re-evaluation of a safety case?</li><li>What are minimum requirements for data sources, and **data quality**, in relation to training of the system? What about algorithm/system calibration?</li><li>How are sources of **uncertainty** handled?</li><li>How are **ethical considerations** embedded in our algorithms?</li></ul> | <ul><li>[ISO/IEC JTC 1/SC 42][18], including:<ul><li>ISO 5259-x Data Quality</li><li>ISO 5392 - AI Ref. Architecture</li><li>ISO AWI 8200 - Controllability of automated AI systems</li><li>ISO 5469 - Functional Safety and AI</li><li>ISO 23894 - AI Risk Management</li><li>ISO 24027 - Bias in AI</li></ul></li><li>[ISO/AWI PAS 8800][19]</li></ul> |
| **Consumer Ratings and Testing Protocols**<br><br>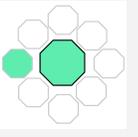 | <ul><li>What's the **minimum set of tests** an ADS should be trialed on?</li><li>What types of **ratings systems** would the public feel more attuned to and should thus be employed to garner **public trust** and acceptance? Should different ratings be established for safety versus other consumer-valued qualifications?</li><li>Can **tests and protocols devised for L2 and/or L3** systems be expanded/ augmented or otherwise applied to L4 systems?</li><li>How can testing protocols be appropriately mapped to various ADS use-cases and associated ODDs? How can we discourage inappropriate comparisons across ADS with diverse capabilities?</li><li>How can tests be mapped to behavioral competencies and how do different intended ODDs impact how each behavioral competency is evaluated?</li></ul> | <ul><li>[ISO/TC 204][20], including:<ul><li>ISO 21737 - Low Speed Testing</li><li>ISO/CD TR 21734-3 - Automated driving bus</li><li>ISO/AWI 23792-2 Motorway chauffeur systems</li><li>ISO 19237:2017 Pedestrian detection and collision</li></ul></li></ul> |
| | <ul><li>How do you **systematically evaluate the soundness** and applicability of proposed protocols?</li><li>How do you **prioritize** the landscape of externally proposed tests?</li></ul> | mitigation systems<br>● [EuroNCAP updates][21]<br>● [US NCAP updates][22]<br>● Testing Annexes in UNECE GRVA ([AEB][23]) |
| **Risk Acceptance and Safety Impact**<br>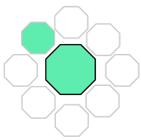 | <ul><li>What type of **risk acceptance criteria** should be used? If multiple, should they be summarized into a single pass/fail criterion?</li><li>What type of **validation targets are acceptable** (comparison to what human)?</li><li>Is it necessary to provide evidence of positive risk balance (PRB)? Is PRB just one acceptance criterion out of many to prove Absence of Unreasonable Risk (AUR)?</li><li>How can you compare performance **across ODDs**?</li><li>How do you compare **predictive**/estimated approaches to **reactive**/observation-based ones?</li><li>What's the minimum set of performance metrics to consider?</li><li>How do you define **preventable vs. unavoidable** collisions</li><li>Should you distinguish between **collisions caused by the ADS** versus those **caused by others**?</li><li>What is the set of outcomes of interest, **beyond collisions**, to evaluate an appropriate safety impact?</li><li>What are appropriate data sources, or, conversely, **data limitations** that affect the setting of appropriate benchmarks?</li><li>What are suitable safety margins for comparison to benchmarks? Should those be set as multipliers and/or added standard deviations?</li><li>When assessing safety benefits, should one constrain</li></ul> | ● ISO 5083 - ADS Design and V&V (ISO/AWI TS 5083)<br>● UL 4600 - Safety of Autonomous Products (UL, 2022)<br>● BSI PAS 1881 - Assuring the operational safety of automated vehicles. Specification (BSI, 2022)<br>● EU Commission Implementing Regulation for ADS Type Approval (EU, 2022)<br>● [French Decree][24]<br>● [UK Law Commission AV Joint Report][25]<br>● [German ADS] |

| | themselves to the strict ODD definition? Non-conforming agents will behave beyond in-scope ODD features (e.g., high speeds) so should extensions of the ODD be considered? |  |
|---|---|---|

## Conclusion

In this paper the authors present a framework that abstracts and organizes the landscape of ADS safety standards into high level and long term themes. The categories presented in the ADS Safety Standards Organization Wheel are then leveraged to develop and align on associated research questions that haven't yet reached full industry consensus. The hope is that framing like the one presented here will allow developers and regulators to better understand and organize ongoing ADS safety standard efforts and the subsequent documentation generated by SDOs. Beyond organizing this area around the question of "how safe is safe enough?," for ADS applications, the framework also allows for the mapping of specific topics of interest to the broader categories presented in the framework, with the potential to identify missing topics and subtopics, including missing research questions.

---

[26] https://bmdv.bund.de/SharedDocs/DE/Pressemitteilungen/2022/008-wissing-verordnung-zum-autonomen-fahren.html